\documentclass[letter]{ieice}
\usepackage[dvips]{graphicx}
\usepackage[fleqn]{amsmath}
\usepackage {amssymb,amsthm,amsbsy}
\usepackage[varg]{txfonts}
\usepackage{enumerate}
\field{A}
\title{On a Class of Almost Difference Sets Constructed by Using the Ding-Helleseth-Martinsen’s Constructions }
\authorlist{
\authorentry[mlqiecully@163.com]{Minglong Qi}{m}{etab1}
\authorentry{Shengwu Xiong}{n}{etab1}
\authorentry{Jingling Yuan}{n}{etab1}
\authorentry{Wenbi Rao}{n}{etab1}
\authorentry{Luo Zhong}{n}{etab1}
}
\affiliate[etab1]{School of Computer Science and Technology, Wuhan University of Technology, Mafangshan West Campus, 430070 Wuhan City, China }

\theoremstyle{definition}

\newtheorem{sec1_thm1}{Theorem}[section]
\newtheorem{sec1_thm2}[sec1_thm1]{Theorem}
\newtheorem{sec1_cor1}{Corollary}[section]
\newtheorem{sec1_cor2}[sec1_cor1]{Corollary}

\newtheorem{sec3_remark1}{Remark}[section]

\newtheorem{sec3_thm1}{Theorem}[section]
\newtheorem{sec3_thm2}[sec3_thm1]{Theorem}

\newtheorem{sec3_lemma1}{Lemma}[section]
\newtheorem{sec3_lemma2}[sec3_lemma1]{Lemma}
\newtheorem{sec3_lemma3}[sec3_lemma1]{Lemma}
\newtheorem{sec3_lemma4}[sec3_lemma1]{Lemma}
\newtheorem{sec3_lemma5}[sec3_lemma1]{Lemma}
\newtheorem{sec3_lemma6}[sec3_lemma1]{Lemma}
\newtheorem{sec3_lemma7}[sec3_lemma1]{Lemma}
\begin{document}
\maketitle
\begin{summary}
Pseudorandom binary sequences with optimal balance and autocorrelation have many applications in stream cipher, communication, coding theory, etc. It is known that binary sequences with three-level autocorrelation should have an almost difference set as their characteristic sets. How to construct new families of almost difference set is an important research topic in such fields as communication, coding theory and cryptography. In a work of Ding, Helleseth, and Martinsen in 2001, the authors developed a new method, known as the Ding-Helleseth-Martinsen’s Constructions in literature, of constructing an almost difference set from product sets of $\mathit{GF}(2) $ and the union of two cyclotomic classes of order four. In the present paper, we have constructed two classes of almost difference set with product sets between $ \mathit{GF}(2) $ and union sets of the cyclotomic classes of order 12 using that method. In addition, we could find there do not exist  the Ding-Helleseth-Martinsen’s Constructions for the cyclotomic classes of order six, eight and ten.
\end{summary}
\begin{keywords}
Binary sequence, three-level autocorrelation,the Ding-Helleseth-Martinsen’s Constructions, almost difference set, cyclotomic classes of order twelve.
\end{keywords}

\section{Introduction}
Let $ (A,+) $ be an Abelian group with $ n $ elements and $ D $ be a $ k$-subset of $ A $. Define the distance function $ d_{D}(e)=|(D+e)\cap D| $, where $ D+e=\lbrace x+e\,|\,x\in D \ \text{and}\  e\in D\setminus \lbrace 0\rbrace\rbrace$. $ D $ is referred to as an $ (n,\,k,\,\lambda,\,t) $ almost difference set if $ d_{D}(e) $ takes on the value $ \lambda $ altogether $ t $ times and on the value $ \lambda+1 $ altogether $ n-1-t $ times when $ e $  ranges over all the nonzero elements of $ A $. Binary sequences with three-level autocorrelation can be constructed from almost difference sets as their characteristic sets \cite{B1x,B2x}. Especially, almost difference sets with parameters $ (n,\,(n-1)/2,\,(n-5)/4,\,(n-1)/4) $ give  corresponding binary sequences with optimal three-level autocorrelation and optimal balance among $ 0 $'s and $ 1 $'s \cite{B1x}. In that sense, constructing optimal binary sequences is equivalent to find corresponding almost difference sets. $ D $ is called a $ (n,\;k,\;\lambda) $ difference set if the distance function  $ d_{D}(e) $ takes on the value $ \lambda $ altogether $ n-1 $ times  when $ e $  ranges over all the nonzero elements of $ A $.

There are several ways of construction of almost difference set. We cite a few of those, for example, generalized cyclotomy can be used to construct almost difference set \cite{B3x}, the so-called Lempel-Cohn-Eastman’s Construction uses set exponents and cyclotomy for the construction of almost difference set \cite{B4x,B5x}. Davis found almost difference sets with following parameters \cite{B6x}:
\begin{enumerate}[(1)]
\item $ (4\cdot 3^{2n},\;2(3^{2n}-3^{n}),\;3^{2n}-2\cdot3^{n},\;3^{2n}-1) $ in $ \mathit{H} \times \mathbf{Z}^{2}_{3^{n}} $, where $ \mathit{H} $ is a group of order four;
\item $ ((q+1)q^{2},\;q(q+1),\;q,\;q^{2}-1) $ in  $ \mathit{H} \times \mathbf{EA}(q^{2}) $, where $ \mathbf{EA}(q^{2}) $ denotes the additive group $ (\mathit{GF}(q^{2}),\;+) $ and $ \mathit{H} $ is a group of order $ q+1 $.
\end{enumerate}
Perfect nonlinear functions (PN) have not only many applications in cryptography \cite{B8x,B9x}, optimal constant-composition codes, and signal sets \cite{B10x,B11x}, but also can be used to construct difference sets \cite{B7x}, and almost difference sets \cite{B2x}.  Almost difference sets can be constructed from difference sets as well, for instance see \cite{B2x,B12x} .

Let $ q=df+1 $ be a power of an odd prime, $ \alpha $ be a primitive element of extension field $ \mathit{GF}(q) $. Define the cosets $ \mathit{D}^{(d,q)}_{i}=\lbrace\alpha^{kd+i}\,|\,0\leq k<f,\ 0\leq i<d\rbrace $, which are called the cyclotomic classes of order $ d $ with respect to $ \mathit{GF}(q) $. It is obvious that $\mathit{GF}(q)^{*}=\bigcup_{i=0}^{d-1} \mathit{D}^{(d,q)}_{i} $. The constants $ (m,\;n)_{d}=|(\mathit{D}^{(d,q)}_{m}+1)\cap \mathit{D}^{(d,q)}_{n}| $ are known as the cyclotomic numbers of order $ d $ with respect to $ \mathit{GF}(q) $.
It is possible to construct almost difference sets with cyclotomic classes of given order, for instance, see \cite{B1x,B2x,B13x,B14x}.

In the present paper, we are concerned with the so-called  the Ding-Helleseth-Martinsen’s Constructions of almost difference set using the cyclotomic classes of order four \cite{B2x,B14x}, and apply the method into the cyclotomic classes of order twelve. By computer investigation, we found that no almost difference sets could be constructed from the cyclotomic classes of order six, eight and ten using that method. However, two classes of almost difference sets can be obtained by using  the Ding-Helleseth-Martinsen’s Constructions with the cyclotomic classes of order twelve. Let $ q\equiv 5\pmod 8 $, then $ q $ can be expanded into $ q=s^{2}+4t^{2} $ with $ s\equiv 1 \pmod 4 $ \cite{B15x}. The Ding-Helleseth-Martinsen’s Constructions can be outlined by the following two theorems:
\begin{sec1_thm1}[Theorem 5 in \cite{B2x}, see also \cite{B14x}]\label{lab-sec1-thn1}
Let $ i,j,l\in \lbrace 0,\,1,\,2\,,3\rbrace $ be three pairwise distinct integers, and define 
\begin{equation*}
\mathit{C}=\lbrace 0\rbrace\times\left(\mathit{D}^{(4,q)}_{i}\cup \mathit{D}^{(4,q)}_{j}\right)\cup \lbrace 1\rbrace\times\left(\mathit{D}^{(4,q)}_{l}\cup \mathit{D}^{(4,q)}_{j}\right).
\end{equation*}
Then $ \mathit{C} $ is an $ (n,\,(n-2)/2,\,(n-6)/4,\,(3n-6)/4) $ almost difference set of $ \mathit{A}=\mathit{GF}(2)\times \mathit{GF}(q) $ if
\begin{enumerate}[(1)]
\item $ t=1 $ and $ (i,\,j,\,l)= (0,\,1,\,3)$ or $ (0,\,2,\,1) $, or 
\item $ s=1 $ and $ (i,\,j,\,l)= (1,\,0,\,3)$ or $ (0,\,1,\,2) $.
\end{enumerate}
\end{sec1_thm1} 
\begin{sec1_thm2}[Theorem 6 in \cite{B2x}, see also \cite{B14x}]\label{lab-sec1-thn2}
Let $ i,j,l\in \lbrace 0,\,1,\,2\,,3\rbrace $ be three pairwise distinct integers, and define 
\begin{equation*}
\mathit{C}=\lbrace 0\rbrace\times\left(\mathit{D}^{(4,q)}_{i}\cup \mathit{D}^{(4,q)}_{j}\right)\cup \lbrace 1\rbrace\times\left(\mathit{D}^{(4,q)}_{l}\cup \mathit{D}^{(4,q)}_{j}\right) \cup \lbrace 0,\;0\rbrace.
\end{equation*}
Then $ \mathit{C} $ is an $ (n,\,n/2,\,(n-2)/4,\,(3n-2)/4) $ almost difference set of $ \mathit{A}=\mathit{GF}(2)\times \mathit{GF}(q) $ if
\begin{enumerate}[(1)]
\item $ t=1 $ and 
\begin{equation*}
(i,\,j,\,l)\in \lbrace (0,\,1,\,3),\,(0,\,2,\,3),\, (1,\,2,\,0),\,(1,\,3,\,0)\rbrace
\end{equation*}
or 
\item $ s=1 $ and 
\begin{equation*}
 (i,\,j,\,l)\in \lbrace (0,\,1,\,2),\,(0,\,3,\,2),\, (1,\,0,\,3),\,(1,\,2,\,3)\rbrace.
\end{equation*}
\end{enumerate}
\end{sec1_thm2} 

The next corollary extends Theorem \ref{lab-sec1-thn1} of which the proof method can be found in \cite[eq.(2), eq.(3), and the proof of Lemma4 and Theorem 1]{B14x}.
\begin{sec1_cor1}\label{lab-sec1-cor1}
Let $ i,j,l\in \lbrace 0,\,1,\,2,\,3\rbrace $ be three pairwise distinct integers, and define 
\begin{equation*}
\mathit{C}=\lbrace 0\rbrace\times\left(\mathit{D}^{(4,q)}_{i}\cup \mathit{D}^{(4,q)}_{j}\right)\cup \lbrace 1\rbrace\times\left(\mathit{D}^{(4,q)}_{l}\cup \mathit{D}^{(4,q)}_{j}\right).
\end{equation*}
Then $ \mathit{C} $ is an $ (n,\,(n-2)/2,\,(n-6)/4,\,(3n-6)/4) $ almost difference set of $ \mathit{A}=\mathit{GF}(2)\times \mathit{GF}(q) $ if
\begin{enumerate}[(1)]
\item $ t=1 $ and 
\begin{equation*}
\begin{split}
 &(i,\,j,\,l)\in \lbrace (0,\,1,\,3),\,(0,\,2,\,1),\,(1,\,2,\,0),\,(1,\,3,\,2),\\
 &(2,\,0,\,3),\,(2,\,3,\,1),\,(3,\,1,\,0),\,(3,\,0,\,2)\rbrace
\end{split}
\end{equation*}
or 
\item $ t=-1 $ and 
\begin{equation*}
\begin{split}
 &(i,\,j,\,l)\in \lbrace (0,\,2,\,3),\,(0,\,3,\,1),\,(1,\,0,\,2),\,(1,\,3,\,0),\\
 &(2,\,0,\,1),\,(2,\,1,\,3),\,(3,\,1,\,2),\,(3,\,2,\,0)\rbrace
\end{split}
\end{equation*}
or 
\item $ s=1 $ and 
\begin{equation*}
\begin{split}
&(i,\,j,\,l)\in \lbrace (0,\,1,\,2),\,(0,\,3,\,2),\,(1,\,0,\,3),\,(1,\,2,\,3),\\
&(2,\,1,\,0),\,(2,\,3,\,0),\,(3,\,0,\,1),\,(3,\,2,\,1)\rbrace,
\end{split}
\end{equation*}
\end{enumerate}
where $ n=|\mathit{GF}(2)\times \mathit{GF}(q)|=2q $.
\end{sec1_cor1}

By a computational research all the triples $ (i,\,j,\,l) $ satisfying Theorem \ref{lab-sec1-thn2} can be found and the extension is outlined in the following corollary, for which the formulae of distance function and the proof method can be found in \cite[eq.(2), eq.(4), and the proof of Lemma4 and Theorem 1]{B14x}.
\begin{sec1_cor2}\label{lab-sec1-cor2}
Let $ i,j,l\in \lbrace 0,\,1,\,2,\,3\rbrace $ be three pairwise distinct integers, and define 
\begin{equation*}
\mathit{C}=\lbrace 0\rbrace\times\left(\mathit{D}^{(4,q)}_{i}\cup \mathit{D}^{(4,q)}_{j}\right)\cup \lbrace 1\rbrace\times\left(\mathit{D}^{(4,q)}_{l}\cup \mathit{D}^{(4,q)}_{j}\right)\cup \lbrace 0,\;0\rbrace.
\end{equation*}
Then $ \mathit{C} $ is an $ (n,\,n/2,\,(n-2)/4,\,(3n-2)/4) $ almost difference set of $ \mathit{A}=\mathit{GF}(2)\times \mathit{GF}(q) $ if
\begin{enumerate}[(1)]
\item $ t=1 $ and 
\begin{equation*}
\begin{split}
  &(i,\,j,\,l)\in \lbrace (0,\,1,\,3),\,(0,\,2,\,3),\,(1,\,2,\,0),\,(1,\,3,\,0),\,\\
                        &(2,\,0,\,1),\,(2,\,3,\,1),\,(3,\,0,\,2),\,(3,\,1,\,2) \rbrace
\end{split}
\end{equation*}
or 
\item $ t=-1 $ and 
\begin{equation*}
\begin{split}
 &(i,\,j,\,l)\in \lbrace (0,\,2,\,1),\,(0,\,3,\,1),\,(1,\,0,\,2),\,(1,\,3,\,2),\,\\
                        &(2,\,0,\,3),\,(2,\,1,\,3),\,(3,\,1,\,0),\,(3,\,2,\,0) \rbrace
\end{split}
\end{equation*}
or 
\item $ s=1 $ and 
\begin{equation*}
\begin{split}
  &(i,\,j,\,l)\in \lbrace (0,\,1,\,2),\,(0,\,3,\,2),\,(1,\,0,\,3),\,(1,\,2,\,3),\,\\
                        &(2,\,1,\,0),\,(2,\,3,\,0),\,(3,\,0,\,1),\,(3,\,2,\,1) \rbrace
\end{split}
\end{equation*}
\end{enumerate}
where $ n=|\mathit{GF}(2)\times \mathit{GF}(q)|=2q $.
\end{sec1_cor2}

 The rest of the present paper is structured as follows: in Section 2, basic concepts and data on the cyclotomic classes of order 12 will be introduced. In section 3 we present the main results of the present paper: two new families of almost difference sets constructed by using the Ding-Helleseth-Martinsen’s Constructions and the cyclotomic classes of order 12. In section 4 a brief concluding remark will be given.
\section{The cyclotomic classes of order twelve}
Let $ p=12f+1 $ be an odd prime. Then, $ p $ can be expressed as $ p=x^{2}+4y^{2} =A^{2}+3B^{2}$, where $ x\equiv 1\pmod 4 $ and $ A\equiv 1\pmod 6 $. In an earlier work of Dickson than Whiteman's \cite{B15x}, Dickson pointed out that the $ 144 $ cyclotomic numbers of order twelve depend solely on the parameters $ p $, $ A,\ B,\ x $, and $ y $. For example, if 2 is a cubic residue of $ p $, 3 is a biquadratic residue of $ p $, and $ f $ is odd, then
\begin{equation*}
144(0,\ 2)_{12}=p+1-2A+24B-12x.
\end{equation*}

In order to define some more generally  necessary concepts, let  $ p=ef+1 $ be an odd prime. Recall that we actually deal with the case $ e=12 $. Let 
$ \beta=\exp(\tfrac{2\pi \vec{j}}{e}) $ be a $ e^{th} $ root of unity 
where $ \vec{j}=\sqrt{-1} $. Let $ a\in Z_{p}^{*} $ where $ Z_{p}^{*} $ denotes the multiplicative group modulo $ p $, and $ Ind(a) $ denote the index of $ a $ modulo $ p $ with respect to a given primitive root. In the theory of cyclotomy, the Jacobi sum plays a fundamental role that can be defined by the following equation for a pair of integers $ m $ and $ n $:
\begin{equation}\label{lab-jacobi-sum-def}
\phi(\beta^{m},\ \beta^{n})=\sum_{a+b\equiv 1\pmod p}\beta^{mInd(a)+nInd(b)},
\end{equation}
where  $ a,\ b\in Z_{p}^{*} $. Another parameter is needed to classify the distinct cyclotomic numbers of order twelve according to $ f $ odd or even, and defined by
\begin{equation}\label{lab-c-parameter-def}
c=\frac{\phi(\beta^{3},\ \beta)}{\phi(\beta^{5},\ \beta)}.
\end{equation}
Let $ g $ be a fixed primitive root modulo $ p $, $ M,\ M^{'}\in  Z_{p}^{*} $ such that $ g^{M}\equiv 2\pmod p $ and  $ g^{M^{'}}\equiv 3\pmod p $. Hereafter, we keep the meanings of the symbols  $ M,\ M^{'}$ and $ c $ defined in (\ref{lab-c-parameter-def}),  unchanged. There are 31 distinct cyclotomic numbers of order twelve which depend on whether $ f $ is odd or not, on what is the residue of $ M^{'}\pmod 4 $, on what is the residue of $ M\pmod 6 $, and on what is the value of the parameter $ c $ defined in (\ref{lab-c-parameter-def}). For $ f $ odd, there are six sets of the 31 distinct cyclotomic numbers of order twelve, they are classified as follows:
\begin{enumerate}[(1)]
\item $ f $ odd, $ M^{'}\equiv 0 \pmod 4,\  M\equiv 1\pmod 6 $ and $ c=\beta^{3} $,
\item $ f $ odd, $ M^{'}\equiv 0 \pmod 4,\  M\equiv 3\pmod 6 $ and $ c=\beta^{3} $,
\item $ f $ odd, $ M^{'}\equiv 2 \pmod 4,\  M\equiv 1\pmod 6 $ and $ c=1 $,
\item $ f $ odd, $ M^{'}\equiv 2 \pmod 4,\  M\equiv 1\pmod 6 $ and $ c=-1 $,
\item $ f $ odd, $ M^{'}\equiv 2 \pmod 4,\  M\equiv 3\pmod 6 $ and $ c=1 $,
\item $ f $ odd, $ M^{'}\equiv 2 \pmod 4,\  M\equiv 3\pmod 6 $ and $ c=-1 $.
\end{enumerate}

Let $ \overrightarrow{Var}=(p,\;A,\;B,\;x,\;y,\;1)^{T} $ denote the column vector for a given odd prime $ p=12f+1 $ where the supscript $ T $ means transpose operation, and $\overrightarrow{CN}_{i} $ represent the  column vector of which the entries consist of the 31 distinct cyclotomic numbers of the above $ i^{th} $ case. Since each of 144 cyclotomic numbers of order twelve can be decomposed into a linear combination of all the entries of $ \overrightarrow{Var} $, there is a 31 by 6 matrix, $ M_{i} $, that links $\overrightarrow{CN}_{i} $ and  $ \overrightarrow{Var} $. In fact, we have $144 \overrightarrow{CN}_{i}= M_{i}\overrightarrow{Var} $ for $ 1\leq i\leq 6 $ \cite{B16x}. Hereafter, the meanings of $ \overrightarrow{Var},\ \overrightarrow{CN}_{i} $ and $  M_{i} $ are kept unchanged. Due to limited space, only $ M_{1} $ is shown in the paper (see Table \ref{lab-table-M1}), for other coefficient matrices from $ M_{2} $ to $ M_{6} $, readers may refer to the original literature of Whiteman \cite{B16x}.

\begin{table}[tb]
\caption{Equalities of $ (h,\ k)_{12} $,  $ f $ odd}
\label{lab-table-cyclotomic-number}
{\renewcommand{\tabcolsep}{0.15cm}
\begin{center}
\begin{tabular}{|c|c|c|c|c|c|c|c|c|c|c|c|c|}
\hline
hk & 0 & 1 & 2 & 3 & 4 & 5 & 6 & 7 & 8 & 9 & 10 & 11 \\
\hline
0 & 00 & 01 & 02 & 03 & 04 & 05 & 06 & 07 & 08 & 09 & 0X & 0Y\\
\hline
1 & 10 & 11 & 12 & 13 & 14 & 15 & 07 & 05 & 15 & 19 & 1X & 1Y\\
\hline
2 & 20 & 21 & 22 & 23 & 24 & 19 & 08 & 15 & 04 & 14 & 24 & 2Y\\
\hline
3 & 30 & 31 & 32 & 30 & 2Y & 1X & 09 & 19 & 14 & 03 & 13 & 23\\
\hline
4 & 22 & 32 & 42 & 31 & 20 & 1Y & 0X & 1X & 24 & 13 & 02 & 12\\
\hline
5 & 11 & 21 & 31 & 32 & 21 & 10 & 0Y & 1Y & 2Y & 23 & 12 & 01\\
\hline
6 & 00 & 10 & 20 & 30 & 22 & 11 & 00 & 10 & 20 & 30 & 22 & 11\\
\hline
7 & 10 & 0Y & 1Y & 2Y & 23 & 12 & 01 & 11 & 21 & 31 & 32 & 21\\
\hline
8 & 20 & 1Y & 0X & 1X & 24 & 13 & 02 & 12 & 22 & 32 & 42 & 31\\
\hline
9 & 30 & 2Y & 1X & 09 & 19 & 14 & 03 & 13 & 23 & 30 & 31 & 32\\
\hline
10 & 22 & 23 & 24 & 19 & 08 & 15 & 04 & 14 & 24 & 2Y & 20 & 21\\
\hline
11 & 11 & 12 & 13 & 14 & 15 & 07 & 05 & 15 & 19 & 1X & 1Y & 10\\
\hline
\end{tabular}
\end{center}
}
\end{table}
\begin{table}[tb]
\caption{$144 \overrightarrow{CN}_{1}= M_{1}\overrightarrow{Var}  $}
\label{lab-table-M1}
{\renewcommand{\tabcolsep}{0.15cm}
\begin{center}
\begin{tabular}{|c|c c c c c c|}
\hline
$ 144(h,\ k)_{12} $ & p & A & B & x & y & 1\\
\hline
$ 144(0,\ 0)_{12} $ & 1 & -6 & 0 & 0 & -16 & -23 \\
$ 144(0,\ 1)_{12} $ & 1 & 4 & 24 & -18 & -24 & -24 \\
$ 144(0,\ 2)_{12} $ & 1 & -2 & -24 & -12 & 0 & -24 \\
$ 144(0,\ 3)_{12} $ & 1 & 18 & 0 & 0 & 32 & 18 \\
$ 144(0,\ 4)_{12} $ & 1 & -12 & 0 & 6 & -16 & -16 \\
$ 144(0,\ 5)_{12} $ & 1 & -2 & -24 & -12 & 0 & -24 \\
$ 144(0,\ 6)_{12} $ & 1 & -14 & 24 & 0 & 48 & -14 \\
$ 144(0,\ 7)_{12} $ & 1 & 12 & 0 & 6 & 8 & 12 \\
$ 144(0,\ 8)_{12} $ & 1 & 6 & 0 & 12 & -16 & -16 \\
$ 144(0,\ 9)_{12} $ & 1 & -14 & 0 & 0 & 0 & -14 \\
$ 144(0,\ 10)_{12} $ & 1 & 4 & 0 & 6 & 0 & 6 \\
$ 144(0,\ 11)_{12} $ & 1 & 6 & 0 & 12 & -16 & -16 \\
$ 144(1,\ 0)_{12} $ & 1 & 0 & 12 & 6 & 8 & -11 \\
$ 144(1,\ 1)_{12} $ & 1 & 6 & 0 & 0 & 8 & -11 \\
$ 144(1,\ 2)_{12} $ & 1 & -12 & 0 & 6 & -16 & -16 \\
$ 144(1,\ 3)_{12} $ & 1 & 4 & -12 & 6 & -24 & -24 \\
$ 144(1,\ 4)_{12} $ & 1 & 6 & 36 & -12 & -16 & -16 \\
$ 144(1,\ 5)_{12} $ & 1 & 0 & -12 & -6 & 8 & -12 \\
$ 144(1,\ 9)_{12} $ & 1 & -12 & 12 & 6 & 8 & -12 \\
$ 144(1,\ 10)_{12} $ & 1 & -6 & 0 & 0 & 8 & -6 \\
$ 144(1,\ 11)_{12} $ & 1 & 4 & 0 & 6 & 0 & 6 \\
$ 144(2,\ 0)_{12} $ & 1 & 6 & 0 & 0 & 8 & -11 \\
$ 144(2,\ 1)_{12} $ & 1 & 0 & -12 & -6 & 8 & -12 \\
$ 144(2,\ 2)_{12} $ & 1 & -12 & 0 & -6 & 8 & -11 \\
$ 144(2,\ 3)_{12} $ & 1 & -6 & 0 & 0 & 8 & -6 \\
$ 144(2,\ 4)_{12} $ & 1 & 12 & 0 & 6 & 8 & 12 \\
$ 144(2,\ 11)_{12} $ & 1 & 0 & -24 & -6 & -16 & -24 \\
$ 144(3,\ 0)_{12} $ & 1 & 6 & -12 & 0 & -16 & -11 \\
$ 144(3,\ 1)_{12} $ & 1 & 0 & 0 & -6 & 32 & -6 \\
$ 144(3,\ 2)_{12} $ & 1 & -2 & 12 & 12 & 0 & -2 \\
$ 144(4,\ 2)_{12} $ & 1 & 4 & 24 & -18 & -24 & -24\\
\hline
\end{tabular}
\end{center}
}
\end{table}

In Table \ref{lab-table-cyclotomic-number}, the relationship between 144 cyclotomic numbers and the 31 distinct irreducible ones for $ f $ odd is demonstrated. An entry $ hk $ with $ 0\leq h\leq 4 $ and $ 0\leq k\leq 9 $ stands for the cyclotomic number $ (h,\ k)_{12} $, an  entry $ hX $ with $ 0\leq h\leq 4 $ denotes the cyclotomic number $ (h,\ 10)_{12} $, and  an  entry $ hY $ with $ 0\leq h\leq 4 $ is equal to the cyclotomic number $ (h,\ 11)_{12} $.

\section{New families of almost difference set constructed  using the Ding-Helleseth-Martinsen’s Constructions and the cyclotomic classes of order 12}
Through the present section the following notation is kept unchanged. Let $ q=12f+1 $
 be a power of an odd prime with $ f $ odd, $ D_{i}^{(12,q)} $ denote the $ i^{th} $ cyclotomic class of order 12 with $ 0\leq i<12 $, $ I,J\subset \mathbf{Z}_{12} $ be index subsets whose entries are pairwise distinct and such that $ |I|,|J|=6 $. Define $ \mathit{D}_{I}=\bigcup_{i\in I}D_{i}^{(12,q)},\  \mathit{D}_{J}=\bigcup_{j\in J}D_{j}^{(12,q)},\ \mathit{C}=\lbrace 0\rbrace\times  \mathit{D}_{I}\cup \lbrace 1\rbrace\times  \mathit{D}_{J}$ and $ ,\mathit{C}^{'}=\lbrace 0\rbrace\times  \mathit{D}_{I}\cup \lbrace 1\rbrace\times  \mathit{D}_{J}\cup \lbrace(0,\,0) \rbrace$. Define also the following distance functions
 \begin{equation*}
 \begin{split}
 d_{I}(w)&=|( \mathit{D}_{I}+w)\cap  \mathit{D}_{I}|,\\
 d_{I,J}(w)&=|( \mathit{D}_{I}+w)\cap  \mathit{D}_{J}|,\\
 d_{\mathit{C}}(w_{1},\,w_{2})&=|\bigl(\mathit{C}+(w_{1},\,w_{2}) \bigr)\cap \mathit{C}|,\\
 d_{\mathit{C}^{'}}(w_{1},\,w_{2})&=|\bigl(\mathit{C}^{'}+(w_{1},\,w_{2}) \bigr)\cap \mathit{C}^{'}|,
 \end{split}
 \end{equation*}
 where $ w,\ w_{2} \in \mathit{GF}(q)\setminus \lbrace 0\rbrace$ and $ w_{1} \in \mathit{GF}(2) $. It is clear that  $ q $ can be expressed as $ q=x^{2}+4y^{2} =A^{2}+3B^{2}$, where $ x\equiv 1\pmod 4 $ and $ A\equiv 1\pmod 6 $ \cite{B15x}. Let 
  \begin{equation*}
  \begin{split}
  \mathit{Q}&= D_{0}^{(12,q)}\cup  D_{2}^{(12,q)}\cup  D_{4}^{(12,q)}\cup  D_{6}^{(12,q)}\cup  D_{8}^{(12,q)}\cup  D_{10}^{(12,q)},\\
  \mathit{QN}&= D_{1}^{(12,q)}\cup  D_{3}^{(12,q)}\cup  D_{5}^{(12,q)}\cup  D_{7}^{(12,q)}\cup  D_{9}^{(12,q)}\cup  D_{11}^{(12,q)}.
  \end{split}
  \end{equation*}
   Hereafter for the proof of lemmas and theorems of this section, we consider only the case where $ f $ is  odd, $ M^{'}\equiv 0 \pmod 4,\  M\equiv 1\pmod 6 $ and $ c=\beta^{3} $, since for other five cases with $ f $ odd (see Section 2), proof process is similar and it leads to the same results.
   
 The distance functions $  d_{\mathit{C}}(w_{1},\,w_{2}) $ and $ d_{\mathit{C}^{'}}(w_{1},\,w_{2}) $ can be explicitly expanded out in $ d_{I}(w_{2})  $, $ d_{J}(w_{2})  $ and $ d_{I,J}(w_{2}) $, stated by the following two lemmas whose proofs can be found in \cite[eq.(2) and eq.(4)]{B14x}
  \begin{sec3_lemma1}\label{lab-sec3-lamma1}
   \begin{equation*}
   d_{\mathit{C}}(w_{1},\,w_{2})=\\
   \begin{cases}
   |\mathit{D}_{I}|+|\mathit{D}_{J}|&\quad\mbox{if}\ w_{1}=0,w_{2}=0\\
   d_{I}(w_{2})+d_{J}(w_{2})&\quad\mbox{if}\ w_{1}=0,w_{2}\ne 0\\
   d_{I,J}(w_{2})+d_{J,I}(w_{2})&\quad\mbox{if}\ w_{1}=1,w_{2}\ne 0\\
   2|\mathit{D}_{I}\cap \mathit{D}_{J}|&\quad\mbox{if}\ w_{1}=1,w_{2}= 0
   \end{cases}
   \end{equation*}
   \end{sec3_lemma1}
 \begin{sec3_lemma2}\label{lab-sec3-lamma2}
  \begin{equation*}
  \begin{split}
   &d_{\mathit{C}^{'}}(w_{1},\,w_{2})=d_{\mathit{C}}(w_{1},\,w_{2})\\
   &+
    \begin{cases}
      |\mathit{D}_{I}\cap \lbrace w_{2},\,-w_{2}\rbrace|&\quad\mbox{if}\ w_{1}=0,w_{2}\ne 0\\
      |\mathit{D}_{J}\cap \lbrace w_{2},\,-w_{2}\rbrace|&\quad\mbox{if}\ w_{1}=1,w_{2}\ne 0\\
     0 &\quad\mbox{otherwise}.
    \end{cases}
  \end{split}
  \end{equation*}
  \end{sec3_lemma2}
  
  The following lemma gives the value of the distance function $ d_{I}(w) $ for a given index set $ I $:
  \begin{sec3_lemma3}\label{lab-sec3-lamma3}
  \begin{enumerate}[(1)] For a given index set $ I $, the distance function $ d_{I}(w) $ can be determined by
  \item Let $ I\in \bigl\lbrace \lbrace 0,1,4,5,8,9\rbrace,\, \lbrace 2,3,6,7,10,11\rbrace \bigr\rbrace $. Then,
  \begin{equation*}
  d_{I}(w)=
  \begin{cases}
  \dfrac{q-2y-3}{4}&\quad \mbox{if}\ w\in \mathit{Q},\\
  \dfrac{q+2y-3}{4}&\quad \mbox{if}\ w\in \mathit{QN}.
  \end{cases}
  \end{equation*}
  \item Let $ I\in \bigl\lbrace \lbrace 0,3,4,7,8,11\rbrace,\, \lbrace 1,2,5,6,9,10\rbrace \bigr\rbrace $. Then,
    \begin{equation*}
    d_{I}(w)=
    \begin{cases}
    \dfrac{q+2y-3}{4}&\quad \mbox{if}\ w\in \mathit{Q},\\
    \dfrac{q-2y-3}{4}&\quad \mbox{if}\ w\in \mathit{QN}.
    \end{cases}
    \end{equation*}
  \item Let $ I= \lbrace 0,2,4,6,8,10\rbrace$. Then,
    \begin{equation*}
    d_{I}(w)=
    \begin{cases}
    \dfrac{q-5}{4}&\quad \mbox{if}\ w\in \mathit{Q},\\
    \dfrac{q-1}{4}&\quad \mbox{if}\ w\in \mathit{QN}.
    \end{cases}
    \end{equation*}
    \item Let $ I= \lbrace 1,3,5,7,9,11\rbrace$. Then,
       \begin{equation*}
       d_{I}(w)=
       \begin{cases}
       \dfrac{q-1}{4}&\quad \mbox{if}\ w\in \mathit{Q},\\
       \dfrac{q-5}{4}&\quad \mbox{if}\ w\in \mathit{QN}.
       \end{cases}
       \end{equation*}
  \end{enumerate}
  \end{sec3_lemma3}
  \begin{proof}
  See the proof for Lemma \ref{lab-sec3-lamma4}.
  \end{proof}
  
  Let
  \begin{equation*}
  \begin{split}
  I,\,J\in &\bigl\lbrace\lbrace 0,1,4,5,8,9\rbrace,\, \lbrace 2,3,6,7,10,11\rbrace,\\
  &\lbrace 0,3,4,7,8,11\rbrace,\, \lbrace 1,2,5,6,9,10\rbrace \bigr\rbrace,
  \end{split}
  \end{equation*}
  $ I\ne J $ and $ |I\cap J|=3 $. Then, $ d_{I,J}(w) $ and $ d_{J,I}(w) $ can be expressed as functions only in $ q $ and $ x $, as the next lemma asserts:
  \begin{sec3_lemma4}\label{lab-sec3-lamma4}
  Let $ I=\lbrace 0,1,4,5,8,9\rbrace $ and $ J= \lbrace 0,3,4,7,8,11\rbrace$. Then,
  \begin{equation*}
  d_{I,J}(w)=\\
  \begin{cases}
  \frac{q+x-2}{4}&\quad  \mbox{if}\ w\in \mathit{Q},\\
  \frac{q-x-4}{4}&\quad  \mbox{if}\ w\in D_{1}^{(12,q)}\cup D_{5}^{(12,q)}\cup D_{9}^{(12,q)},\\
  \frac{q-x}{4}&\quad  \mbox{if}\ w\in D_{3}^{(12,q)}\cup D_{7}^{(12,q)}\cup D_{11}^{(12,q)},
  \end{cases}
  \end{equation*}
  and 
   \begin{equation*}
    d_{J,I}(w)=\\
    \begin{cases}
    \frac{q+x-2}{4}&\quad  \mbox{if}\ w\in \mathit{Q},\\
    \frac{q-x-4}{4}&\quad  \mbox{if}\ w\in \in D_{3}^{(12,q)}\cup D_{7}^{(12,q)}\cup D_{11}^{(12,q)},\\
    \frac{q-x}{4}&\quad  \mbox{if}\ w\in D_{1}^{(12,q)}\cup D_{5}^{(12,q)}\cup D_{9}^{(12,q)}.
    \end{cases}
    \end{equation*}
  \end{sec3_lemma4}
  \begin{proof}
  We only prove $ d_{I,J}(w) $. Let $ w\in \mathit{GF}(q)\setminus \lbrace 0\rbrace $ and $ w^{-1}\in D_{h}^{(12,q)} $. Remark that $ (i+h,\,j+h)_{12}=\bigl((i+h)\pmod {12},\,(j+h)\pmod {12}\bigr)_{12} $.
  \begin{equation}\label{lab-sec3-lma4-prf01}
  \begin{split}
  d_{I,J}(w)&=|( \mathit{D}_{I}+w)\cap  \mathit{D}_{J}|\\
  &=|(\bigcup_{i\in I}D_{i}^{(12,q)}+w)\cap (\bigcup_{j\in J}D_{j}^{(12,q)})|\\
  &=|\bigl(\bigcup_{i\in I}(D_{i}^{(12,q)}+w)\bigr)\cap (\bigcup_{j\in J}D_{j}^{(12,q)})|\\
  &=|\bigl(\bigcup_{i\in I}(w^{-1}D_{i}^{(12,q)}+1)\bigr)\cap \bigcup_{j\in J}w^{-1}D_{j}^{(12,q)}|\\
   &=|\bigl(\bigcup_{i\in I}(D_{i+h}^{(12,q)}+1)\bigr)\cap (\bigcup_{j\in J}D_{j+h}^{(12,q)})|\\
 &=\sum_{i\in I}\sum_{j\in J}|(D_{i+h}^{(12,q)}+1)\cap D_{j+h}^{(12,q)}|\\
 &=\sum_{i\in I}\sum_{j\in J}(i+h,\,j+h)_{12}.
 \end{split}
  \end{equation}
  Let $ v_{1},v_{2},\cdots,v_{31} $ represent the 31 irreducible and distinct cyclotomic numbers of order 12 (see Table \ref{lab-table-M1}), then, from the last equation of eq.(\ref{lab-sec3-lma4-prf01}) and using Table \ref{lab-table-cyclotomic-number}, we can obtain
  \begin{enumerate}[(1)]
  \item $ D_{h}^{(12,q)}\subset \mathit{Q} $:
  \begin{equation}\label{lab-sec3-lma4-prf02}
  \begin{split}
   d_{I,J}(w)&=v_{1}+v_{10}+v_{12}+v_{13}+v_{14}+2v_{15}+\\
             &2v_{16}+v_{17}+v_{18}+v_{19}+v_{2}+2v_{20}+\\
             &2v_{21}+2v_{22}+v_{23}+2v_{24}+v_{25}+\\
             &2v_{26}+v_{27}+v_{28}+2v_{29}+2v_{30}+\\
             &v_{4}+v_{5}+v_{6}+v_{8}+v_{9}.
  \end{split}
  \end{equation}
   \item $ D_{h}^{(12,q)}\subset D_{1}^{(12,q)}\cup D_{5}^{(12,q)}\cup D_{9}^{(12,q)} $:
    \begin{equation}\label{lab-sec3-lma4-prf03}
     \begin{split}
    d_{I,J}(w)&=v_{1}+3v_{13}+3v_{14}+3v_{17}+3v_{18}+\\
              &3v_{19}+2v_{22}+3v_{23}+2v_{24}+3v_{25}+\\
              &2v_{26}+3v_{27}+3v_{28}+v5+v9. 
     \end{split}
    \end{equation}
   \item $ D_{h}^{(12,q)}\subset D_{3}^{(12,q)}\cup D_{7}^{(12,q)}\cup D_{11}^{(12,q)} $:
      \begin{equation}\label{lab-sec3-lma4-prf04}
       \begin{split}
      d_{I,J}(w)&=v_{10}+3v_{11}+v_{12}+v_{13}+v_{14}+\\
                &2v_{15}+2v_{16}+v_{17}+v_{18}+v_{19}+v_{2}+\\
                &2v_{20}+2v_{21}+v_{23}+v_{25}+v_{27}+v_{28}+\\
                &2v_{29}+3v_{3}+2v_{30}+2v_{31}+v_{4}+\\
                &v_{6}+v_{7}+v_{8}.
       \end{split}
      \end{equation}
  \end{enumerate}
  From Table \ref{lab-table-M1}, we have 
  \begin{equation}\label{lab-sec3-lma4-proof}
  \begin{split}
  v_{1}&=(0,\,0)_{12}=\dfrac{q-6A-16y-23}{144},\\
  v_{2}&=(0,\,1)_{12}=\dfrac{q+4A+24B-18x-24y-24}{144},\\
  v_{3}&=(0,\,2)_{12}=\dfrac{q-2A-24B-12x-24}{144},\\
  &\cdots \cdots\\
  v_{30}&=(3,\,2)_{12}=\dfrac{q-2A+12B+12x-2}{144},\\ 
  v_{31}&=(4,\,2)_{12}=\dfrac{q+4A+24B-18x-24y-24}{144}.
  \end{split}
  \end{equation}
  Substitute the formulae of $ v_{1},\,v_{2},\cdots,v_{31} $ in eq.(\ref{lab-sec3-lma4-proof}) for $ v_{1},\,v_{2},\cdots,v_{31} $ in eq.(\ref{lab-sec3-lma4-prf02})-eq.(\ref{lab-sec3-lma4-prf04}).
  \end{proof}

For two index sets $ I,\ J $ one of which is $ \lbrace 0,2,4,6,8,10\rbrace $ or $ \lbrace 1,3,5,7,9,11\rbrace $, the distance function $  d_{I,J}(w) $ is quite different. Let 
\begin{equation*}
\begin{split}
I,J\in &\bigl\lbrace \lbrace 0, 3, 4, 7, 8, 11\rbrace,\,\lbrace 1, 2, 5, 6, 9, 10\rbrace,\,\lbrace 0,2,4,6,8,10\rbrace \bigr\rbrace\\
&\mbox{or}\\
I,J\in &\bigl\lbrace \lbrace 0, 1, 4, 5, 8, 9\rbrace,\,\lbrace 2, 3, 6, 7, 10, 11\rbrace,\,\lbrace 1, 3, 5, 7, 9, 11\rbrace \bigr\rbrace,
\end{split}
\end{equation*}
and $ |I\cap J|=3 $. Then, the distance function $  d_{I,J}(w) $ is a function only in $ q $, $ x $ and $ y $.
\begin{sec3_lemma5}\label{lab-sec3-lamma5}
 Let $ I=\lbrace 0, 3, 4, 7, 8, 11\rbrace $ and $ J= \lbrace 0,2,4,6,8,10\rbrace$. Then,
  \begin{equation*}
  d_{I,J}(w)=\\
  \begin{cases}
  \frac{q-x-2y-2}{4}&\quad  \mbox{if}\ w\in D_{0}^{(12,q)}\cup D_{4}^{(12,q)}\cup D_{8}^{(12,q)},\\
   \frac{q+x+2y-4}{4}&\quad  \mbox{if}\ w\in D_{2}^{(12,q)}\cup D_{6}^{(12,q)}\cup D_{10}^{(12,q)},\\
   \frac{q+x-2y}{4}&\quad  \mbox{if}\ w\in D_{1}^{(12,q)}\cup D_{5}^{(12,q)}\cup D_{9}^{(12,q)},\\
   \frac{q-x+2y-2}{4}&\quad  \mbox{if}\ w\in D_{3}^{(12,q)}\cup D_{7}^{(12,q)}\cup D_{11}^{(12,q)},
  \end{cases}
  \end{equation*}
  and 
   \begin{equation*}
    d_{J,I}(w)=\\
    \begin{cases}
    \frac{q-x-2y-2}{4}&\quad  \mbox{if}\ w\in D_{2}^{(12,q)}\cup D_{6}^{(12,q)}\cup D_{10}^{(12,q)},\\
     \frac{q+x+2y-4}{4}&\quad  \mbox{if}\ w\in D_{0}^{(12,q)}\cup D_{4}^{(12,q)}\cup D_{8}^{(12,q)},\\
     \frac{q+x-2y}{4}&\quad  \mbox{if}\ w\in D_{3}^{(12,q)}\cup D_{7}^{(12,q)}\cup D_{11}^{(12,q)},\\
     \frac{q-x+2y-2}{4}&\quad  \mbox{if}\ w\in D_{1}^{(12,q)}\cup D_{5}^{(12,q)}\cup D_{9}^{(12,q)}.
    \end{cases}
    \end{equation*}
\end{sec3_lemma5}
\begin{proof}
Similar to Lemma \ref{lab-sec3-lamma4} and omitted due to limited space.
\end{proof}

We need another similar lemma to prove the main theorems.
 Let 
\begin{equation*}
\begin{split}
I,J\in &\bigl\lbrace \lbrace 0, 1, 4, 5, 8, 9\rbrace,\,\lbrace 2, 3, 6, 7, 10, 11\rbrace,\,\lbrace 0,2,4,6,8,10\rbrace \bigr\rbrace\\
&\mbox{or}\\
I,J\in &\bigl\lbrace \lbrace 0, 3, 4, 7, 8, 11\rbrace,\,\lbrace 1, 2, 5, 6, 9, 10\rbrace,\,\lbrace 1, 3, 5, 7, 9, 11\rbrace \bigr\rbrace,
\end{split}
\end{equation*}
and $ |I\cap J|=3 $. Then, the distance function $  d_{I,J}(w) $ is a function only in $ q $, $ x $ and $ y $ as well.
\begin{sec3_lemma6}\label{lab-sec3-lamma6}
 Let $ I=\lbrace 0, 1, 4, 5, 8, 9\rbrace $ and $ J= \lbrace 0,2,4,6,8,10\rbrace$. Then,
  \begin{equation*}
  d_{I,J}(w)=\\
  \begin{cases}
  \frac{q-x+2y-2}{4}&\quad  \mbox{if}\ w\in D_{0}^{(12,q)}\cup D_{4}^{(12,q)}\cup D_{8}^{(12,q)},\\
   \frac{q+x-2y-4}{4}&\quad  \mbox{if}\ w\in D_{2}^{(12,q)}\cup D_{6}^{(12,q)}\cup D_{10}^{(12,q)},\\
   \frac{q+x+2y}{4}&\quad  \mbox{if}\ w\in D_{3}^{(12,q)}\cup D_{7}^{(12,q)}\cup D_{11}^{(12,q)},\\
   \frac{q-x-2y-2}{4}&\quad  \mbox{if}\ w\in D_{1}^{(12,q)}\cup D_{5}^{(12,q)}\cup D_{9}^{(12,q)},
  \end{cases}
  \end{equation*}
  and 
   \begin{equation*}
    d_{J,I}(w)=\\
    \begin{cases}
   \frac{q-x+2y-2}{4}&\quad  \mbox{if}\ w\in D_{2}^{(12,q)}\cup D_{6}^{(12,q)}\cup D_{10}^{(12,q)},\\
      \frac{q+x-2y-4}{4}&\quad  \mbox{if}\ w\in D_{0}^{(12,q)}\cup D_{4}^{(12,q)}\cup D_{8}^{(12,q)},\\
      \frac{q+x+2y}{4}&\quad  \mbox{if}\ w\in D_{1}^{(12,q)}\cup D_{5}^{(12,q)}\cup D_{9}^{(12,q)},\\
      \frac{q-x-2y-2}{4}&\quad  \mbox{if}\ w\in D_{3}^{(12,q)}\cup D_{7}^{(12,q)}\cup D_{11}^{(12,q)}.
    \end{cases}
    \end{equation*}
\end{sec3_lemma6}
\begin{proof}
Similar to Lemma \ref{lab-sec3-lamma4} and omitted due to limited space.
\end{proof}

Now, we are ready to state and prove the first theorem.
\begin{sec3_thm1}\label{lab-sec3-thm1}
Let $ \mathit{C}=\lbrace 0\rbrace\times  \mathit{D}_{I}\cup \lbrace 1\rbrace\times  \mathit{D}_{J} $. Then, $  \mathit{C} $ is  an $ (n,\,(n-2)/2,\,(n-6)/4,\,(3n-6)/4) $ almost difference set of $ \mathit{A}=\mathit{GF}(2)\times \mathit{GF}(q) $ if
\begin{enumerate}[(1)]
\item $ x=1 $ and 
\begin{equation*}
\begin{split}
I,\,J\in&\bigl\lbrace \lbrace 0, 1, 4, 5, 8, 9\rbrace,\,\lbrace 0, 3, 4, 7, 8, 11\rbrace,\\
&\lbrace 1, 2, 5, 6, 9, 10\rbrace,\,\lbrace 2, 3, 6, 7, 10, 11\rbrace\bigr\rbrace
\end{split}
\end{equation*}
such that $ |I\cap J|=3 $, or
\item $ y=1 $ and
\begin{equation*}
I,\,J\in\bigl\lbrace \lbrace 0, 1, 4, 5, 8, 9\rbrace,\,\lbrace 2, 3, 6, 7, 10, 11\rbrace,\,\lbrace1,3,5,7,9,11\rbrace\bigr\rbrace
\end{equation*}
such that $ |I\cap J|=3 $, or
\item $ y=1 $ and
\begin{equation*}
I,\,J\in\bigl\lbrace\lbrace 0, 3, 4, 7, 8, 11\rbrace,\,\lbrace 1, 2, 5, 6, 9, 10\rbrace,\,\lbrace 0, 2, 4, 6, 8, 10\rbrace\bigr\rbrace
\end{equation*}
such that $ |I\cap J|=3 $, or
\item $ y=-1 $ and
\begin{equation*}
I,\,J\in\bigl\lbrace \lbrace 0, 1, 4, 5, 8, 9\rbrace,\,\lbrace 2, 3, 6, 7, 10, 11\rbrace,\,\lbrace 0,2,4,6,8,10\rbrace\bigr\rbrace
\end{equation*}
such that $ |I\cap J|=3 $, or
\item $ y=-1 $ and
\begin{equation*}
I,\,J\in\bigl\lbrace\lbrace 0, 3, 4, 7, 8, 11\rbrace,\,\lbrace 1, 2, 5, 6, 9, 10\rbrace,\,\lbrace 1, 3, 5, 7, 9, 11\rbrace\bigr\rbrace
\end{equation*}
such that $ |I\cap J|=3 $,
\end{enumerate}
where $ n=|\mathit{GF}(2)\times \mathit{GF}(q)|=2q $.
\end{sec3_thm1}
\begin{proof}
We only prove case (1) with $ I=\bigl\lbrace 0, 1, 4, 5, 8, 9\bigr\rbrace $ and $ J=\bigl\lbrace 0, 3, 4, 7, 8, 11 \bigr\rbrace $ since proof for other cases is similar. By Lemma \ref{lab-sec3-lamma1} for $ w_{1}=0 $ and $ w_{2}\ne 0 $,
\begin{equation}\label{lab-sec3-thm1-prf01}
 d_{\mathit{C}}(w_{1},\,w_{2})=d_{I}(w_{2})+d_{J}(w_{2}).
\end{equation}
From Lemma \ref{lab-sec3-lamma3} (1) and (2),  substitute the values of $ d_{I}(w_{2}) $, $ d_{J}(w_{2}) $ into eq.(\ref{lab-sec3-thm1-prf01}), we can obtain
\begin{equation}\label{lab-sec3-thm1-prf-dis1}
d_{\mathit{C}}(w_{1},\,w_{2})=\frac{q-3}{2}\quad \mbox{for}\ w_{2}\in \mathit{Q}\cup \mathit{QN}.
\end{equation} 

By Lemma \ref{lab-sec3-lamma1} for $ w_{1}=1 $ and $ w_{2}\ne 0 $,
\begin{equation}\label{lab-sec3-thm1-prf03}
 d_{\mathit{C}}(w_{1},\,w_{2})=d_{I,J}(w_{2})+d_{J,I}(w_{2}).
\end{equation}
Combining eq.(\ref{lab-sec3-thm1-prf03}) and Lemma \ref{lab-sec3-lamma4}, we get 
\begin{equation}\label{lab-sec3-thm1-prf-dis2}
 d_{\mathit{C}}(w_{1},\,w_{2})=\\
 \begin{cases}
 \frac{q+x-2}{2}&\quad \mbox{if}\ w_{2}\in \mathit{Q},\\
 \frac{q-x-2}{2}&\quad \mbox{if}\ w_{2}\in \mathit{QN}.
 \end{cases}
\end{equation}
Set x=1 into eq.(\ref{lab-sec3-thm1-prf-dis2}), it results to
 \begin{equation}\label{lab-sec3-thm1-prf-dis3}
 d_{\mathit{C}}(w_{1},\,w_{2})=\\
 \begin{cases}
 \frac{q-1}{2}&\quad \mbox{if}\ w_{2}\in \mathit{Q},\\
 \frac{q-3}{2}&\quad \mbox{if}\ w_{2}\in \mathit{QN}.
 \end{cases}
 \end{equation}
 
 Now consider the case  $ w_{1}=1 $ and $ w_{2}= 0 $ in Lamma \ref{lab-sec3-lamma1}.
 \begin{equation}\label{lab-sec3-thm1-prf-dis4}
 \begin{split}
 d_{\mathit{C}}(w_{1},\,w_{2})&=2|\mathit{D}_{I}\cap \mathit{D}_{J}|\\
 &=2|D_{0}^{(12,q)}\cup D_{4}^{(12,q)}\cup D_{8}^{(12,q)}|\\
 &=2\cdot\frac{q-1}{12}\cdot 3\\
 &=\frac{q-1}{2}.
 \end{split}
 \end{equation}
 Combining eq.(\ref{lab-sec3-thm1-prf-dis1}), eq.(\ref{lab-sec3-thm1-prf-dis3}) and eq.(\ref{lab-sec3-thm1-prf-dis4}), we conclude that $ \mathit{C} $ is an $ (2q,\,q-1,\,\frac{q-3}{2},\,\frac{3(q-1)}{2}) $ almost difference set with respect to $ \mathit{GF}(2)\times \mathit{GF}(q) $.
\end{proof}

In Lemma \ref{lab-sec3-lamma2} two extra-terms $\delta_{I}= |\mathit{D}_{I}\cap \lbrace w_{2},\,-w_{2}\rbrace| $ and $ \delta_{J}= |\mathit{D}_{J}\cap \lbrace w_{2},\,-w_{2}\rbrace| $ due to the presence of the element $ (0,\,0) $ in the product set $ \mathit{C}^{'}=\lbrace 0\rbrace\times  \mathit{D}_{I}\cup \lbrace 1\rbrace\times  \mathit{D}_{J}\cup \lbrace(0,\,0) \rbrace $   can be explicitly computed for a given pair of distinct $ I $ and $ J $ by the following lemma:
\begin{sec3_lemma7}\label{lab-sec3-lamma7}
If $ D_{0}^{(12,q)}\cup D_{6}^{(12,q)}\subset \mathit{D}_{I} $ then $ \delta_{I}= |\mathit{D}_{I}\cap \lbrace w_{2},\,-w_{2}\rbrace|=2 $, else if  $ D_{0}^{(12,q)}\subset \mathit{D}_{I} $ or  $  D_{6}^{(12,q)}\subset \mathit{D}_{I} $ then $ \delta_{I}= |\mathit{D}_{I}\cap \lbrace w_{2},\,-w_{2}\rbrace|=1 $, else $ \delta_{I}= |\mathit{D}_{I}\cap \lbrace w_{2},\,-w_{2}\rbrace|=0 $.
\end{sec3_lemma7}
\begin{proof}
Let $ w_{2}^{-1}\in D_{h}^{(12,q)} $. Recall that $ D_{h+i}^{(12,q)}=D_{(h+i)\pmod{12}}^{(12,q)} $. Since $ q=12f+1 $ with $ f $ being odd, $ -1\in D_{6}^{(12,q)} $.
\begin{equation*}
\begin{split}
\delta_{I}&= |\mathit{D}_{I}\cap \lbrace w_{2},\,-w_{2}\rbrace|\\
&=|(\bigcup_{i\in I}D_{i}^{(12,q)})\cap \lbrace w_{2},\,-w_{2}\rbrace|\\
&=|(\bigcup_{i\in I} w_{2}^{-1}D_{i}^{(12,q)})\cap \lbrace1,\,-1\rbrace|\\
&=|(\bigcup_{i\in I} D_{i+h}^{(12,q)})\cap \lbrace1,\,-1\rbrace|\\
&= |\mathit{D}_{I+h}\cap \lbrace 1,\,-1\rbrace|,
\end{split}
\end{equation*}
where $ I+h=\lbrace i+h\,|\,i\in I\rbrace $.
\end{proof}
\begin{sec3_remark1}\label{lab-sec3-remark1}
Let $ I=\bigl\lbrace 0,1,4,5,8,9\bigr\rbrace $ and $ J=\bigl \lbrace 0,3,4,7,8,11\bigr\rbrace$. Then, for all $ w_{2}\in \mathit{GF}(q)\setminus \lbrace 0\rbrace $
\begin{equation}\label{lab-extra-term}
\begin{split}
\delta_{I}&= |\mathit{D}_{I}\cap \lbrace w_{2},\,-w_{2}\rbrace|=1,\\
\delta_{J}&= |\mathit{D}_{J}\cap \lbrace w_{2},\,-w_{2}\rbrace|=1.
\end{split}
\end{equation}
\end{sec3_remark1}

Now we are ready to state and prove the second theorem:
\begin{sec3_thm2}\label{lab-sec3-thm2}
Let $ \mathit{C}^{'}=\lbrace 0\rbrace\times  \mathit{D}_{I}\cup \lbrace 1\rbrace\times  \mathit{D}_{J}\cup \lbrace(0,\,0) \rbrace $. Then, $ \mathit{C}^{'} $ is  an  $ (n,\,n/2,\,(n-2)/4,\,(3n-2)/4) $ almost difference set of $ \mathit{A}=\mathit{GF}(2)\times \mathit{GF}(q) $ if
\begin{enumerate}[(1)]
\item $ x=1 $ and 
\begin{equation*}
\begin{split}
I,\,J\in&\bigl\lbrace \lbrace 0, 1, 4, 5, 8, 9\rbrace,\,\lbrace 0, 3, 4, 7, 8, 11\rbrace,\\
&\lbrace 1, 2, 5, 6, 9, 10\rbrace,\,\lbrace 2, 3, 6, 7, 10, 11\rbrace\bigr\rbrace
\end{split}
\end{equation*}
such that $ |I\cap J|=3 $, or
\item $ y=1 $ and
\begin{equation*}
I,\,J\in\bigl\lbrace \lbrace 0, 1, 4, 5, 8, 9\rbrace,\,\lbrace 2, 3, 6, 7, 10, 11\rbrace,\,\lbrace1,3,5,7,9,11\rbrace\bigr\rbrace
\end{equation*}
such that $ |I\cap J|=3 $, or
\item $ y=1 $ and
\begin{equation*}
I,\,J\in\bigl\lbrace\lbrace 0, 3, 4, 7, 8, 11\rbrace,\,\lbrace 1, 2, 5, 6, 9, 10\rbrace,\,\lbrace 0, 2, 4, 6, 8, 10\rbrace\bigr\rbrace
\end{equation*}
such that $ |I\cap J|=3 $, or
\item $ y=-1 $ and
\begin{equation*}
I,\,J\in\bigl\lbrace \lbrace 0, 1, 4, 5, 8, 9\rbrace,\,\lbrace 2, 3, 6, 7, 10, 11\rbrace,\,\lbrace 0,2,4,6,8,10\rbrace\bigr\rbrace
\end{equation*}
such that $ |I\cap J|=3 $, or
\item $ y=-1 $ and
\begin{equation*}
I,\,J\in\bigl\lbrace\lbrace 0, 3, 4, 7, 8, 11\rbrace,\,\lbrace 1, 2, 5, 6, 9, 10\rbrace,\,\lbrace 1, 3, 5, 7, 9, 11\rbrace\bigr\rbrace
\end{equation*}
such that $ |I\cap J|=3 $,
\end{enumerate}
where $ n=|\mathit{GF}(2)\times \mathit{GF}(q)|=2q $.
\end{sec3_thm2}
\begin{proof}
We only prove case (1) with $ I=\bigl\lbrace 0, 1, 4, 5, 8, 9\bigr\rbrace $ and $ J=\bigl\lbrace 0, 3, 4, 7, 8, 11 \bigr\rbrace $ since proof for other cases is similar.
\begin{enumerate}[(1)]
\item Case $ w_{1}=0 $ and $ w_{2}\ne 0 $. By Lemma \ref{lab-sec3-lamma2}, eq.(\ref{lab-sec3-thm1-prf-dis1}), and Remark \ref{lab-sec3-remark1} eq.(\ref{lab-extra-term}), for all $  w_{2}\in \mathit{GF}(q)\setminus \lbrace 0\rbrace $
\begin{equation}\label{lab-sec3-thm2-dis1}
\begin{split}
 d_{\mathit{C}^{'}}(w_{1},\,w_{2})&= d_{\mathit{C}}(w_{1},\,w_{2})+\delta_{I}\\
 &=\frac{q-3}{2}+1\\
 &=\frac{q-1}{2}. 
 \end{split}
\end{equation}
\item Case $ w_{1}=1 $ and $ w_{2}\ne 0 $. By Lemma \ref{lab-sec3-lamma2}, eq.(\ref{lab-sec3-thm1-prf-dis2}), and Remark \ref{lab-sec3-remark1} eq.(\ref{lab-extra-term}),
\begin{equation}\label{lab-sec3-thm2-dis2}
\begin{split}
 &d_{\mathit{C}^{'}}(w_{1},\,w_{2})=d_{\mathit{C}}(w_{1},\,w_{2})+\delta_{J}\\
 &=
 \begin{cases}
  \frac{q+x-2}{2}+1&\quad \mbox{if}\ w_{2}\in \mathit{Q},\\
  \frac{q-x-2}{2}+1&\quad \mbox{if}\ w_{2}\in \mathit{QN}.
  \end{cases}
\end{split}
\end{equation}
Set $ x=1 $ into eq.(\ref{lab-sec3-thm2-dis2}), it leads to
\begin{equation}\label{lab-sec3-thm2-dis2x}
\begin{split}
 &d_{\mathit{C}^{'}}(w_{1},\,w_{2})= d_{\mathit{C}}(w_{1},\,w_{2})+\delta_{J}\\
 &=
 \begin{cases}
  \frac{q+1}{2}&\quad \mbox{if}\ w_{2}\in \mathit{Q},\\
  \frac{q-1}{2}&\quad \mbox{if}\ w_{2}\in \mathit{QN}.
  \end{cases}
\end{split}
\end{equation}
\item
Case $ w_{1}=1 $ and $ w_{2}= 0 $. By Lemma \ref{lab-sec3-lamma2}, eq.(\ref{lab-sec3-thm1-prf-dis4}),
 \begin{equation}\label{lab-sec3-thm2-dis3}
 \begin{split}
 d_{\mathit{C}^{'}}(w_{1},\,w_{2})&=2|\mathit{D}_{I}\cap \mathit{D}_{J}|\\
 &=2|D_{0}^{(12,q)}\cup D_{4}^{(12,q)}\cup D_{8}^{(12,q)}|\\
 &=2\cdot\frac{q-1}{12}\cdot 3\\
 &=\frac{q-1}{2}.
 \end{split}
 \end{equation}
\end{enumerate}
From eq.(\ref{lab-sec3-thm2-dis1}), eq.(\ref{lab-sec3-thm2-dis2x}) and eq.(\ref{lab-sec3-thm2-dis3}), it is clear that $ \mathit{C}^{'}=\lbrace 0\rbrace\times  \mathit{D}_{I}\cup \lbrace 1\rbrace\times  \mathit{D}_{J}\cup \lbrace(0,\,0) \rbrace $ is an $ (2q,\,q,\,\frac{q-1}{2},\,\frac{3q-1}{2}) $ almost difference set with respect to $ \mathit{A}=\mathit{GF}(2)\times \mathit{GF}(q) $ with  $ n=|\mathit{GF}(2)\times \mathit{GF}(q)|=2q $.
\end{proof}
\section{Conclusion}
The Ding-Helleseth-Martinsen's Constructions is an efficient method to find and construct new almost difference set. In the present paper, we have constructed two classes of almost difference set with product sets between $ \mathit{GF}(2) $ and union sets of the cyclotomic classes of order 12 using that method. Constructing binary sequences with optimal three-level autocorrelation and optimal balance is equivalent to find corresponding almost difference sets as their support sets. It is possible to extend the Ding-Helleseth-Martinsen’s Constructions by using $ \mathit{GF}(4) $ instead of $ \mathit{GF}(2) $. In addition, by computer investigation we found that no the Ding-Helleseth-Martinsen’s Constructions exists for cyclotomic classes of order six, eight and ten.

\end{document}